\documentclass[preprint,12pt]{elsarticle}

\usepackage{amsmath}
\usepackage{amssymb}
\journal{the arXiv.org}

\begin{document}

\begin{frontmatter}

\title{Frequency spectrum of  nonlinear  oscillations and resonance phenomena for graphene plates}

\author{Yuriy Ostapov}

\address{National Technical University of Ukraine "Igor Sikorsky Kyiv Polytechnic Institute", 37 Prosp.Peremohy,  Kyiv, Ukraine 03056.
E-mail: yugo.ost@gmail.com }

\begin{abstract}
The paper studies  oscillations of graphene plates  under the hypothesis that  deformations are much more than the thickness of plate.
In this  most realistic  case the oscillations are described by the system of nonlinear partial differential equations (the F\"{o}ppl-von K\'{a}rm\'{a}n equations).
This system  is reduced to  one nonlinear ordinary differential equation and investigated by means of the Bogoliubov-Mitropolsky asymptotic methods.
As a result, we have  the  real  frequency spectrum for rectangular graphene plates. 
Next we have examined the nonlinear resonance effects under forced oscillations.
These outcomes can apply for variable strain-induced pseudomagnetic fields.
Such fields permit better to understand the properties of flexural phonons connected with transport process.
Resonance phenomena play a leading role in application of graphene plates in engineering constructions.\\


\end{abstract}

\end{frontmatter}

\section{Introduction}

A major direction  of solid state physics is the investigation of graphene (Kittel \cite {kit} Chap.18). There is a need to develop fundamental  analysis of graphene properties on the base of quantum and classical physics. 
These problems are examined in the comprehensive monograph of Katsnelson \cite {kats} as well as in the books of Gorbar \& Sharapov \cite {shar}, Mikhailov \cite {mikh}, Shafraniuk \cite {shaf}. 

In particularly, in the ninth chapter of Katsnelson \cite {kats}  the deformations of graphene plates are considered in the view of the {\it theory elasticity} and  described by the {\it F\"{o}ppl-von K\'{a}rm\'{a}n equations}, which are sufficiently difficult for analysis. 
In addition, several works  are devoted  to the mechanical properties  of graphene plates: Booth et al. \cite {both}; Lee et al. \cite {lee}; Los, Fasolino \& Katsnelson \cite {los}.

 In our paper the solution of the   F\"{o}ppl-von K\'{a}rm\'{a}n equations is proposed on the base of {\it nonlinear oscillations theory}.
 We shall  consider  the nonlinear oscillations of defined mode presented in terms of the product of sinusoidal functions.
Therefore, the stress function is described in the explicit form, and  the system of nonlinear partial differential equations is reduced to  one nonlinear ordinary differential equation.
 The latter is investigated by means of the {\it Bogoliubov-Mitropolsky  asymptotic methods} \cite {bog}. 
 As a result, we have the real {\it frequency spectrum} for   graphene plates under the situation when  deformations are much more than the thickness of plate.
Next we considered the  nonlinear forced oscillations and resonance phenomena.

 In the tenth chapter of Katsnelson \cite {kats} the   strain-induced {\it pseudomagnetic fields} are examined. 
 The review of Amorim et al.\cite {amor}  contains the all-round description of these subjects. 
 Our results concerning the F\"{o}ppl-von K\'{a}rm\'{a}n equations can help to  study the properties of strained graphene.

 The oscillations of graphene plates lead  to the concept of {\it flexural phonons}. The properties of flexural phonons are connected with transport processes (see Katsnelson \cite {kats} Sec.11.4). 
The flexural phonons are considered as well in the papers of Stauber et al. \cite {sta}; Morozov et al. \cite {mor}; Mariani \& von Oppen \cite {mar}; Castro et al. \cite {cast}; Ochoa et al. \cite {och}.
These researches are based on the linear theory of oscillations. 

Besides, the investigation of  oscillations has the important practical meaning.
Graphene plates in engineering constructions  undergo perturbations and  will vibrate.
This can lead to violations of work or even to the destruction of plate.

\section{F\"{o}ppl-von K\'{a}rm\'{a}n equations }

 In this section we consider the equations of oscillations for graphene plates. These equations are based on the elasticity theory \footnote {See Landau \& Lifshitz \cite {lan}.}. 
 The coordinate axes $x, y$ of a rectangular plate are directed along sides. We shall use such denotations:
 $E$ is  {\it Young's modulus}, $\sigma$ is the {\it Poisson ratio}, $\rho$ is the mass density, $h$ is  the thickness of plate, 
 \begin{equation}
  \Delta  = \frac {{\partial}^2} {{\partial}x^2 } + \frac {{\partial}^2 } {{\partial}y^2 } ,                                                           
\end{equation}
 
 \begin{equation}
   D =  \frac{Eh^3 } {12 (1-{\sigma}^2)} .                                                         
\end{equation}
 Let $\chi (x,y) $ be the  stress function, $\zeta (x,y)$ be the out-of-plane deflection of the plate. 
Then  we get the next equations to describe the nonlinear oscillations of graphene plates \footnote {See Landau \& Lifshitz \cite {lan} \S 14.}: 
 
 \begin{equation}
D {\Delta}^2 \zeta - h L (\zeta,\chi) + h\rho \frac {{\partial}^2 \zeta} {{\partial}t^2 } = 0,
 \end{equation}
 
 \begin{equation}
 {\Delta}^2 \chi + \frac {E} {2} L (\zeta,\zeta) = 0, 
 \end{equation}
where  
 \begin{equation}
   L (\zeta,\chi) = \frac {{\partial}^2 \zeta} {{\partial}x^2 } \frac {{\partial}^2 \chi} {{\partial}y^2 }  +  \frac {{\partial}^2 \zeta} {{\partial}y^2 } \frac {{\partial}^2 \chi} {{\partial}x^2 } - 2 \frac {{\partial}^2 \zeta} {{\partial}x {\partial}y} \frac {{\partial}^2 \chi} {{\partial}x{\partial}y }.                                                        
\end{equation}
 
  To avoid cumbersome calculations, we shall assume in the following that   plate edges rest on a fixed support (Landau \& Lifshitz \cite {lan} \S 12).

\section{Linear oscillations}
 
Preparatory to analysing nonlinear oscillations, first we shall consider linear oscillations. Such vibrations will be under the hypothesis of small enough deformations.
 Then we derive the linear equation \footnote {See Landau \& Lifshitz \cite {lan} \S 25.}:

\begin{equation}
D {\Delta}^2 \zeta  + h\rho \frac {{\partial}^2 \zeta} {{\partial}t^2 } = 0.
 \end{equation}

Taking into account  the boundary conditions, we present the solution of Eq.(6)  in terms of the series

\begin{equation}
 \zeta (x,y,t) = \sum _{n,m} A_{mn}(t) sin (\frac {m\pi x}{a}) sin (\frac {n\pi y}{b}),
 \end{equation}
where $a$ is the length of plate and $b$ is the width of plate, $m$ and $n$ are positive integers.
Each member of the series corresponds to the defined mode of oscillations.
Substituting the series (7) in Eq.(6), multiplying by $sin (\frac {m\pi x}{a}) sin (\frac {n\pi y}{b})$ and integrating over  all area of plate,  we obtain the equation

\begin{equation}
 \frac {d^2 A_{mn}} {dt^2 } +  {\omega}_{0,mn}^2 A_{mn} = 0,
 \end{equation}
  where ${\omega}_{0,mn}$ are the frequencies of linear oscillations that are computed using the expression
  
 \begin{equation}
{\omega}_{0,mn}^2  = \frac {Eh^2{\pi}^4 } {12 \rho (1 - {\sigma}^2)} {\bigg[{(\frac {m} {a}) }^2 + {(\frac {n} {b}) }^2 \bigg ]}^2.
 \end{equation} 
 
 Consequently, we have found the frequency spectrum of linear oscillations for graphene plates.
 
 Consider now a plane wave as the solution of Eq.(6). Substituting 
 
\begin{equation}
\zeta = c \cdot exp \,[ i \,(k_{x} x + k_{y} y - {\omega}_{0,mn} t)]
 \end{equation} 
in Eq.(6), one obtains

\begin{equation}
- \rho {\omega}_{0,mn}^2 + \frac {D}{h} k^4 =0.
 \end{equation} 

 Then
 
 \begin{equation}
    k^2 = {\pi}^2\bigg[{(\frac {m} {a}) }^2 + {(\frac {n} {b}) }^2 \bigg ].
 \end{equation} 
 
 Since arbitrary oscillations are described by the superposition of plane waves (10), 
the method of secondary quantization leads to the concept of flexural phonons with the energy $ {\cal E} = \hbar{\omega}_{0,mn}$ and
the quasi-momentum $\vec {p} = \hbar \vec{k}$ (see Kosevich \cite {kos} Sec.6.6.; Landau \& Lifshitz \cite {land} \S \S 71-72).

\section{Nonlinear oscillations of presented mode}

Let us consider the nonlinear oscillations described by Eqs.(3) and (4).
We shall study the nonlinear oscillations of defined mode(for the fixed values $m$ and $n$):

\begin{equation}
    \zeta = f_{mn}(t) sin (\frac {m\pi x}{a}) sin (\frac {n\pi y}{b}).
 \end{equation} 

Substituting this expression in

\begin{equation}
    L (\zeta,\zeta) = 2 \frac {{\partial}^2\zeta}{\partial x^2}	\frac {{\partial}^2\zeta}{\partial y^2} -  2{(\frac {{\partial}^2\zeta}{\partial x \partial y})}^2,
 \end{equation} 
one obtains

\begin{equation}
    L (\zeta,\zeta) = 2{f^{2}_{mn}} {\pi}^4 \frac {m^2n^2}{a^2b^2} \bigg [sin ^2 (\frac {m\pi x}{a}) sin ^2 (\frac {n\pi y}{b}) - cos ^2 (\frac {m\pi x}{a}) cos ^2 (\frac {n\pi y}{b}) \bigg]
 \end{equation} 

 As a result, we have the equation
 
 \begin{equation}
 {\Delta}^2 \chi = \frac {1}{2} E{f^{2}_{mn}} {\pi}^4 \frac {m^2n^2}{a^2b^2} \bigg [cos  \frac {2m\pi x}{a}  + cos  \frac {2n\pi y}{b} \bigg].
 \end{equation}

 We present the solution of this equation in terms of the expression
 
\begin{equation}
 \chi =  A \cdot cos  \frac {2m\pi x}{a}  + B \cdot cos  \frac {2n\pi y}{b}.
 \end{equation} 
 
 Then we shall find that
 
 \begin{equation}
 A = \frac {Ef^{2}_{mn}} {32} \frac {n^2a^2} {m^2 b^2} , \; B = \frac {Ef^{2}_{mn}} {32} \frac {m^2b^2} {n^2 a^2} . 
 \end{equation}
 
Substituting the expressions (13) and (17) in Eq.(3), one derives

\begin{eqnarray}
& \frac {D} {h} {\Delta}^2 \zeta  -  (\frac {{\partial}^2\chi}{\partial y^2} \frac {{\partial}^2\zeta}{\partial x^2} +  \frac {{\partial}^2\chi}{\partial x^2} \frac {{\partial}^2\zeta}{\partial y^2} - 
 2 \frac {{\partial}^2 \zeta} {{\partial}x {\partial}y} \frac {{\partial}^2 \chi} {{\partial}x{\partial}y }) +  \rho \frac {{\partial}^2 \zeta} {{\partial}t^2 }=\nonumber \\
& = [f_{mn} \frac {D} {h} {\pi}^4 (\frac {m^4}{a^4} + \frac {n^4}{b^4}  + 2 \frac {m^2 n^2}{a^2 b^2}) - B {(\frac {2\pi n} {b})}^2 f_{mn} \frac {m^2 {\pi}^2}{a^2} cos \frac {2\pi n y} {b}   -  \nonumber \\
& - A {(\frac {2\pi m} {a})}^2 f_{mn} \frac {n^2 {\pi}^2}{b^2} cos \frac {2\pi m x} {a} + \rho \frac {d^2 f_{mn}} {dt^2 }] sin \frac {m \pi x} {a} sin \frac {n \pi y} {b} = 0.
 \end{eqnarray} 
  
Using the values of A and B from the relations (18 ), Eq.(19) is transformed to

 \begin{eqnarray}
&  [f_{mn} \frac {D} {h} {\pi}^4 {(\frac {m^2}{a^2} + \frac {n^2}{b^2} )}^2 -  \frac {Ef^{3}_{mn} {\pi}^4} {8} (\frac {m^4} {a^4} cos \frac {2\pi n y} {b}   +   \frac {n^4} {b^4}  cos \frac {2\pi m x} {a}) + \nonumber \\
& +\rho \frac {d^2 f_{mn}} {dt^2 }] sin \frac {m \pi x} {a} sin \frac {n \pi y} {b} = 0.
 \end{eqnarray} 
 
 Multiplying by $sin (\frac {m\pi x}{a}) sin (\frac {n\pi y}{b})$ and integrating over  all area of plate,  we  shall find that
 
\begin{equation}
  f_{mn} \frac {D} {h} {\pi}^4 {(\frac {m^2}{a^2} + \frac {n^2}{b^2} )}^2 +   \frac {Ef^{3}_{mn}} {16} {\pi}^4 (\frac {m^4} {a^4}    +   \frac {n^4} {b^4}  ) + \rho \frac {d^2 f_{mn}} {dt^2 } = 0,
 \end{equation} 
since 

\begin{eqnarray}
& \int_{0}^{a} {sin}^2 (\frac {m \pi x}{a})dx  \int_{0}^{b} {sin}^2 (\frac {n \pi y}{b}) dy  =  \frac {ab}{4} , \nonumber \\
& \int_{0}^{a} {sin}^2 (\frac {m \pi x}{a})cos (\frac {2\pi m x} {a})dx  \int_{0}^{b} {sin}^2 (\frac {n \pi y}{b}) dy  =  -\frac {ab}{8} , \nonumber \\
 & \int_{0}^{a} {sin}^2 ( \frac {m \pi x}{a})dx  \int_{0}^{b} {sin}^2 (\frac {n \pi y}{b}) cos (\frac {2\pi n y} {b})dy  =  -\frac {ab}{8}. 
\end{eqnarray}
  
We shall rewrite Eq.(21) as 

\begin{equation}
 \frac {d^2 f_{mn}} {dt^2 } + {\omega}^{2}_{0,mn}f_{mn} + K f^{3}_{mn}  = 0,
 \end{equation} 
where 

\begin{equation}
 K = \frac {E {\pi}^4} {16 \rho}(\frac {m^4} {a^4} + \frac {n^4} {b^4}  ).
  \end{equation}
  
  Thus, we have obtained the nonlinear ordinary differential equation.

To study Eq.(23), we shall use the asymptotic methods proposed by Bogoliubov-Mitropolsky \cite {bog}. 
These methods  are explained briefly in Appendix A. 
 Applying general formulas from Appendix A for  Eq.(23), we have $u = f_{mn}, \, f_{mn} = \alpha cos \, \psi$,
  
  \begin{equation}
  \varepsilon \phi (u) = - K f^{3}_{mn}, \; {\omega}^2={\omega}^{2}_{0,mn},
 \end{equation}

 \begin{equation}
  C_{1} = - \frac {K {\alpha}^3}{\pi\varepsilon} \int_{0}^{2\pi} {cos}^4 \psi d\psi = - \frac {3K{\alpha}^3}{4\varepsilon}.
 \end{equation} 
 
For the corrected frequency we have the formula

\begin{equation}
   {\omega}_{I,mn}(\alpha)= {\omega}_{0,mn} + \frac {3K {\alpha}^2}{8 {\omega}_{0,mn}}
 \end{equation} 
It should be noted that we have found the frequency spectrum for the defined mode of nonlinear oscillations.
There are as well  {\it anharmonic effects} connected with combination frequencies (Landau \& Lifshitz \cite {lan} \S 26).
However, it is difficult enough  to realize corresponding calculations.

Since in the first approximation the oscillations of plates are harmonic,  the plane wave

\begin{equation}
\zeta = c \cdot exp \, [i \,(k_{x} x + k_{y} y - \omega_{I,mn} (\alpha) t)]
 \end{equation} 
 describes the nonlinear oscillations of defined mode. Then one can speak about the flexural phonons with the energy ${\cal E} = \hbar{\omega}_{I,mn}(\alpha)$ and
the quasi-momentum $\vec {p} = \hbar \vec{k}$ in  assuming that

\begin{equation}
k^2 = \sqrt { \frac {h\rho}{D}} {\omega}_{I,mn}(\alpha).  
 \end{equation}

 \section{Linear forced oscillations}
 
 Consider the equation 
 
 \begin{equation}
D {\Delta}^2 \zeta  + h\rho \frac {{\partial}^2 \zeta} {{\partial}t^2 } = q (x,\,y) sin \,\Omega t - 2\rho h \beta \frac {{\partial} \zeta} {{\partial}t },
 \end{equation}
where $q (x,\,y)$ is the intensity of transverse loading, the latter member describes the dissipation force. 
Substituting the expression (7) in Eq.(57), multiplying by $sin (\frac {m\pi x}{a}) sin (\frac {n\pi y}{b})$ and integrating over  all area of plate,  we obtain the equation

\begin{equation}
 \frac {d^2 A_{mn}} {dt^2 }+ 2 \beta\frac {d A_{mn}} {dt } +  {\omega}_{0,mn}^2 A_{mn} = Q_{mn}sin \,\Omega t ,
 \end{equation}
where

\begin{equation}
 Q_{mn} = \frac {4}{abh \rho }\int_{0}^{a} \int_{0}^{b} q (x,\,y)sin (\frac {m\pi x}{a}) sin (\frac {n\pi y}{b}) dx \,dy
 \end{equation}

Presenting the solution of Eq.(58) in terms of

\begin{equation}
 A_{mn} = B_{mn}cos (\,{\Omega} t + \theta),
 \end{equation}
 
we find that 

\begin{equation}
 B_{mn} = \frac {  Q_{mn}} {\sqrt {    {   ({\omega}_{0,mn}^2 - {\Omega}^2 )   }^2  + 4 {\beta}^2  {\Omega}^2    }  }.
 \end{equation}
 
 In the case of the resonance $\Omega = {\omega}_{0,mn}$ we have

\begin{equation}
 B_{mn} = \frac {  Q_{mn}} {  2 \beta  \Omega  }.
 \end{equation} 
 
 It should be borne in mind that under the resonance $\Omega = {\omega}_{0,mn}$ the only oscillationsmode $sin (\frac {m\pi x}{a}) sin (\frac {n\pi y}{b})$
 is selected  in the expression (7) since $B_{mn} > > B_{ij}$ under the small enough $\beta$.
 
 \section{Nonlinear resonance phenomena}

 In the case of forced oscillations we get the equation
 
 \begin{equation}
D {\Delta}^2 \zeta - h L (\zeta,\chi) + h\rho \frac {{\partial}^2 \zeta} {{\partial}t^2 } = q (x,\,y) sin \,\Omega t - 2\rho h \beta \frac {{\partial} \zeta} {{\partial}t }
 \end{equation}
  instead of Eq.(3).  Repeating the above reasoning from the section 4, we obtain the equation 
 
\begin{equation}
 \frac {d^2 f_{mn}} {dt^2 } + 2 \beta\frac {d f_{mn}} {dt } + {\omega}^{2}_{0,mn}f_{mn} + K f^{3}_{mn}  =  Q_{mn}sin \,\Omega t
 \end{equation}  
 
 To study Eq.(37), we use the {\it equivalent linearization} that is considered in Appendix B.

 Then  $u = f_{mn}, \, f_{mn} = \alpha cos \, \psi$,
 
 \begin{equation}
  \varepsilon \phi (u, \frac {du} {dt}) = - K f^{3}_{mn}  - 2\beta \frac {df^{mn}} {dt}, \; {\omega}^2={\omega}^{2}_{0,mn}, R = Q_{mn}.
 \end{equation} 
 
 Now we can compute $ {\lambda}_{e}(\alpha)$ and $ k_{e}(\alpha)$:
 
 \begin{equation}
     {\lambda}_{e}(\alpha) =  \frac {1} {\pi \alpha{\omega}_{0,mn}} \int_{0}^{2\pi} (- K{\alpha}^3 {cos}^3 \, \psi+ 2\alpha \beta \Omega  sin \, \psi) sin \, \psi d \psi = 2 \beta,
 \end{equation}
 
\begin{equation}
   k_{e}(\alpha)  =  {\omega}^2_{0,mn} - \frac {1} {\pi \alpha} \int_{0}^{2\pi} (- K{\alpha}^3 {cos}^3 \, \psi + 2\alpha \beta \Omega  sin \, \psi) cos \, \psi d \psi = {\omega}^2_{0,mn} +\frac {3K {\alpha}^2}{4}.
 \end{equation} 
 
 Then ${\delta}_{e}(\alpha)=  \beta$, and

 \begin{equation}
   {\omega}^2_{e}  = {\omega}^2_{0,mn} +\frac {3K {\alpha}^2}{4}.
 \end{equation}

The relation between the amplitude of stationary oscillations $\alpha$ and the frequency of external force $\Omega$  gets the mode

\begin{equation}
 {\alpha}^2 = \frac { Q_{mn}^2} {   {({\omega}_{0,mn}^2 +\frac {3K {\alpha}^2}{4} - {\Omega}^2 )   }^2  + 4 {\beta}^2  {\Omega}^2 } . 
 \end{equation}

\section{Conclusion}

We have considered the F\"{o}ppl-von K\'{a}rm\'{a}n  equations  to describe the nonlinear oscillations of graphene plates.
Using the presentation of oscillation mode  as the product of sinusoidal functions, we have reduced  these equations to one nonlinear ordinary differential equation.
As a result we have found the real frequency spectrum for  graphene plates. 
Next we have studied the nonlinear resonance phenomena in the case of forced oscillations.

An important outcome of our analysis is that the investigations of nonlinear oscillations of graphene plates can be applied  for variable strain-induced pseudomagnetic fields.
Such fields permit better to understand the properties of flexural phonons connected with transport processes.

\appendix

\section{Asymptotic methods}

Consider the differential equation with small parameter $\varepsilon$:

\begin{equation}
 \frac {d^2 u} {dt^2 } + {\omega}^{2}u =  \varepsilon \phi (u,\frac {du} {dt}).
 \end{equation} 
 
We shall present the solution of this equation in terms of the sum \footnote {See Bogoliubov \& Mitropolsky \cite {bog} \S 1.}:

\begin{equation}
  u = \alpha cos \, \psi + \varepsilon u_{1} (\alpha, \psi) + {\varepsilon}^{2} u_{2} (\alpha, \psi) + ...
 \end{equation} 
where $u_{1} (\alpha, \psi), \, u_{2} (\alpha, \psi), ... $ are the periodic functions of the angle $\psi$ with the period $2\pi$, $\alpha$ and $\psi$ are the time functions determined as

\begin{equation}
 \frac {d \alpha} {dt } = \varepsilon A_{1}(\alpha) + {\varepsilon}^2 A_{2}(\alpha) + ... ,
 \end{equation}
 
 \begin{equation}
 \frac {d \psi} {dt } =    \omega + \varepsilon B_{1}(\alpha) + {\varepsilon}^2 B_{2}(\alpha) + ...
 \end{equation}

 At the beginning, we consider the first approximation:

 \begin{equation}
u= \alpha cos \,\psi,
 \end{equation}

\begin{equation}
 \frac {d \alpha} {dt } = \varepsilon A_{1}(\alpha), 
 \end{equation}
 
 \begin{equation}
 \frac {d \psi} {dt } =    \omega + \varepsilon B_{1}(\alpha), 
 \end{equation}
 where

 \begin{equation}
  A_{1}(\alpha) = - \frac {1} {2\pi \omega} \int_{0}^{2\pi} \phi (\alpha cos \, \psi, -\alpha\omega sin \, \psi) sin \, \psi d \psi ,
 \end{equation}

  \begin{equation}
  B_{1}(\alpha) = - \frac {1} {2\pi\alpha  \omega} \int_{0}^{2\pi} \phi (\alpha cos \,\psi, -\alpha\omega sin \, \psi) cos \, \psi d \psi. 
 \end{equation}

 In particularly, consider the equation of specific interest for the free nonlinear oscillations of plates \footnote {See Bogoliubov \& Mitropolsky \cite {bog} \S 2.}:

\begin{equation}
 \frac {d^2 u} {dt^2 } + {\omega}^{2}u =  \varepsilon \phi (u).
 \end{equation}
 
Since $\phi (\alpha cos \, \psi)$ is the even function, we have the expansion

\begin{equation}
 \phi (\alpha cos \, \psi) = \sum_{n=0}^{\infty} C_{n} cos \, n \psi,
 \end{equation}
where

\begin{equation}
C_{n} = \frac {1}{\pi} \int_{0}^{2\pi} \phi (\alpha cos \, \psi)cos \, n \psi d\psi.
 \end{equation}
 
Then
\begin{equation}
   A_{1}(\alpha)= 0,  \;   B_{1} (\alpha) = - \frac {C_{1}(\alpha)} {2\omega \alpha},  
 \end{equation}

Thus, in the first approximation:
 
\begin{equation}
u_{I} = \alpha cos \,\psi,
 \end{equation}

\begin{equation}
 \frac {d \alpha} {dt } = 0, 
 \end{equation}
 
 \begin{equation}
 \frac {d \psi} {dt } =    \omega - \frac {\varepsilon C_{1}(\alpha)}{2\omega \,\alpha} = {\omega}_{I}(\alpha).
 \end{equation} 
 
Consequently, the amplitude $\alpha$ does not depend on time and preserves an initial value. The phase $\psi$ is equal to:

\begin{equation}
 \psi = {\omega}_{I}(\alpha)t +\theta.
 \end{equation}
Then  in the first approximation the oscillations are {\it harmonic} with the frequency ${\omega}_{I}(\alpha)$.

\section{Equivalent linearization}

Consider  the equation 
 
 \begin{equation}
 \frac {d^2 u} {dt^2 } + {\omega}^{2}u =  \varepsilon \phi (u,\frac {du} {dt}) + R sin \, \Omega t .
 \end{equation}

Next we shall analyze the resonance $\Omega \approx \omega $. The first approximation has the form  $u = \alpha cos (\Omega t +\theta ) $.
The functions $\alpha (t)$ and $\theta(t) $ satisfy the relations  (see Bogoliubov \& Mitropolsky \cite {bog} \S 15):

\begin{equation}
 \frac {d \alpha} {dt } = - {\delta}_{e}(\alpha) \alpha   - \frac {R \,cos \, \theta} {\omega + \Omega},
 \end{equation}
 
 \begin{equation}
 \frac {d \theta} {dt } =    {\omega}_{e}(\alpha) - \Omega + \frac {R \,sin \, \theta} {\alpha(\omega + \Omega)},
 \end{equation}
 where 

 \begin{equation}
 {\delta}_{e}(\alpha)= \frac {1}{2}  {\lambda}_{e}(\alpha) , \;      {\lambda}_{e}(\alpha) =  \frac {\varepsilon} {\pi \alpha\omega} \int_{0}^{2\pi} \phi (\alpha cos \, \psi, -\alpha\Omega sin \, \psi) sin \, \psi d \psi ,
 \end{equation}

  \begin{equation}
 {\omega}_{e}(\alpha)= \sqrt{  k_{e}(\alpha)} , \;     k_{e}(\alpha)  =  {\omega}^2 - \frac {\varepsilon} {\pi \alpha} \int_{0}^{2\pi} \phi (\alpha cos \, \psi, -\alpha\Omega sin \, \psi) cos \, \psi d \psi ,
 \end{equation}
and $ \psi = \Omega t +\theta$.

 The relation between the amplitude of stationary oscillations $\alpha$ and the frequency of external force $\Omega$  is determined as
 
 \begin{equation}
 {\alpha}^2 = \frac { R^2} {   {({\omega}_{e}^2 - {\Omega}^2 )   }^2  + 4 {\delta}_{e}^2  {\Omega}^2 .           }  
 \end{equation}

\end{document}